\documentclass[twocolumn,aps]{revtex4}

\usepackage{graphicx}
\usepackage{dcolumn}
\usepackage{bm}
\usepackage{indentfirst}

\begin{document}
\title{A triple quantum dot in a single wall carbon nanotube}
\author{K. Grove-Rasmussen$^{a,b)}$}
\email{grove@will.brl.ntt.co.jp}
\author{H. I. J\o rgensen$^{b)}$}
\author{T. Hayashi$^{a)}$}
\author{P. E. Lindelof~$^{b)}$}
\author{T. Fujisawa$^{a)}$}
\affiliation{$^{a)}$NTT Basic Research Laboratories, NTT Corporation, 3-1 Morinosato
Wakamiya, Atsugi-shi, Kanagawa 243-0198, Japan\\
 $^{b)}$Nano-Science Center, Niels Bohr Institute, University of Copenhagen, Universitetsparken 5, 2100~Copenhagen \O , Denmark}
\date{\today}
\begin{abstract}
A top-gated single wall carbon nanotube is used to define three
coupled quantum dots in series between two electrodes. The
additional electron number on each quantum dot is controlled by
top-gate voltages allowing for current measurements of single,
double and triple quantum dot stability diagrams. Simulations using
a capacitor model including tunnel coupling between neighboring dots
captures the observed behavior with good agreement. Furthermore,
anti-crossings between indirectly coupled levels and higher order
cotunneling are discussed.

\end{abstract}

\pacs{}

\maketitle

Carbon nanotubes (CNTs), a promising material for quantum
information devices\cite{Sapmaz2006NanoLett,Loss1998PhRvA}, are
one-dimensional systems with remarkable coherency of electrons. It
is therefore attractive to electrostatically define quantum dots,
also known as artificial atoms, along the length of the CNT.
Electron transport through a two-atomic molecule\cite{vanderWiel}
consisting of two coupled quantum dots in a
CNT\cite{Biercuk2003NanoLett,Mason2004Sci,Biercuk2005NanoLett,Sapmaz2006NanoLett,Graeber2006PhRvB,Graeber2006SeScT,Jorgensen2006ApPhL}
 probes molecular states in
which the electron is delocalized over the two quantum dots. Scaling
the system further up to three coupled quantum
dots\cite{WegewijsPRB,WegewijsJJAP2001} or a tri-atomic artificial
molecule enables the study of more intriguing phenomena related to
electrostatics\cite{Waugh1995PhRvL,Waugh1996PhRvB,SchroerPRB} and
molecular states of the triple quantum dot (quantum superposition of
three levels). Coupled three level systems might allow for future
experiments inspired by the field of quantum
optics\cite{BrandesReview,Renzioni2001Adiabatictransfer,Greentree2004PhRvB,BenakkerEurophys,Emary2007}
and are also attractive from a quantum information point of
view\cite{Meierqbitcluster,Zhang2004entangler}. In contrast to GaAs
defined triple quantum dots, which can be (or sometimes
unintentionally are) arranged in a triangular
 configuration\cite{Gaudreau2006PhRvL,Korkusinski2007,Ihn,StopaPRL2002Rectifying,VidanAPL2004,SagaraTriangularEntangle,Emary2007},
the CNT geometry ensures the serial and simplest configuration,
where electrons only tunnel between neighboring dots. Furthermore,
the more challenging experiments of investigating the serial triple
quantum dot Kondo effect\cite{Zitko2006PhRvB} are yet to be
addressed experimentally.

In this Letter we present measurement on a top-gated single wall
carbon nanotube showing individual control of three coupled quantum
dots defined between the source and drain electrodes. The
characteristics are understood as transport through molecular states
rather than sequential tunneling through three dots. Charging
effects with different capacitive and tunnel couplings between
neighboring quantum dots are investigated by an electrostatic
capacitor model including first order tunneling processes. Finally,
a discussion of triple quantum dot characteristics as second order
anti-crossings and higher order cotunneling is presented.
\begin{figure}[b!]
\begin{center}
\includegraphics[width=0.48\textwidth]{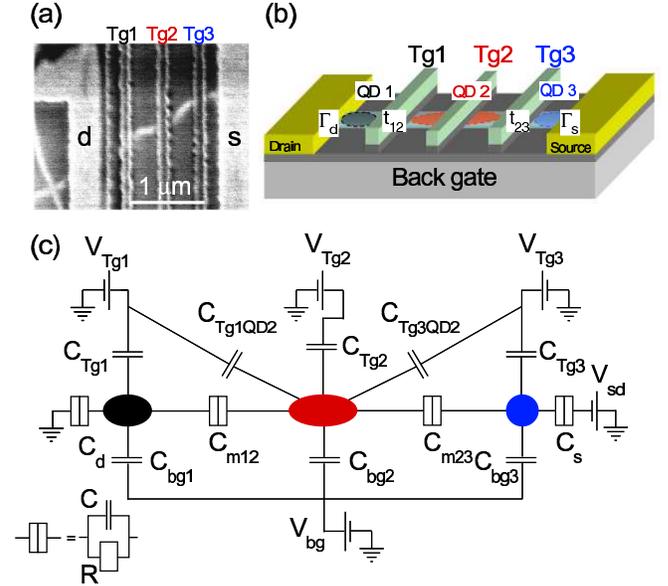}
\end{center}
\caption{(a) Scanning electron micrograph (2\,keV) of the central
part of a similar device with the nanotube lying below three
top-gates. (b) Schematic illustration of the device and the three
serially coupled quantum dots formed due to the top-gates (black,
red and blue ovals). The tunnel rate at the nanotube/metal interfaces is
given by $\Gamma_{d,s}$ and in this particular device tunnel
couplings $t_{12}$ and $t_{23}$ are introduced below Tg1 and Tg3,
respectively. (c) Electrostatic capacitance model including the most
significant capacitances to model the observed behavior (see Fig.\
\ref{figure2} and \ref{fig3}). Tg1/3 tunes the electron number on
QD1/3 and QD2, the latter effect represented by the cross
capacitances. Furthermore, each nanotube quantum dot has capacitive
coupling to the backgate. \label{fig1}}
\end{figure}

The devices are fabricated by an initial step to define Pt/Ti
(40\,nm/5\,nm) alignment marks on top of a SiO$_2$ capped (500\,nm)
highly doped Si substrate. Subsequently, catalyst islands of
Iron-nitrate (Fe(NO$_3$)$_3)$, Molybdenum acetate
(MoO$_2$(CH$_3$COO)$_2$) and Al-oxide nano-sized particles in
methanol are deposited relative to the alignment marks. Single wall
carbon nanotubes are grown from the catalyst islands by chemical
vapor deposition at 850-950\,$^\circ$C in a mixture of methane,
argon and hydrogen. Electrodes (Au/Ti) are defined some microns away
from the catalyst islands in hope that one nanotube bridges the
electrode gap\cite{Jorgensen2006ApPhL}. Top-gate electrodes (8\,nm
Al oxide followed by 10\,nm Al and 25\,nm Ti) are defined on top of
the nanotube and a final optical lithography step is used to make
the bonding pads. A scanning electron micrograph of a similar device
with a nanotube crossing the gap between the two electrodes is shown
in Fig.\ \ref{fig1}(a).

Figure \ref{fig1}(b) schematically shows the cross-section of the
device. Three top-gates are defined (Tg1-3) between the source and
drain electrodes contacting the nanotube. The tunnel rate between
the drain/source electrode and the nanotube is $\Gamma_{d,s}$, which
to some extent is determined by the choice of electrode
material\cite{Javey2003Sm,KimAPL2003,Babic}. The choice of Ti for
this device yields tunnel barriers high enough to obtain single
electron tunneling characteristics (see below).

In case of a semiconducting CNT, the backgate can adjust the
electrochemical potential of the CNT into the valence/conduction band while
the top-gates can introduce barriers below the gates by locally
tuning the electrochemical potential into the band gap. By inducing tunnel
barriers
 under selected top-gate electrodes, several quantum
dot structures can be imagined as {\em e.g.}\ a (tunable) single,
double, triple or even quadruple quantum dot. However, sometimes
barriers are formed at zero gate voltage probably due to defects
introduced during the electron beam lithography or evaporation
process. In this particular device, tunnel barriers are formed under
the two outermost top-gates (Tg1 and Tg3), but not under the central
top-gate (Tg2) as evidenced from the following measurements. Each
top-gate controls the electrostatic potential of (some of) the dots,
QD1, QD2 and QD3, while the tunnel barriers are less affected. The
obtained triple quantum dot is illustrated in Fig.\ \ref{fig1}(b)
with the different colored ovals showing the position of the three
quantum dots (black, red and blue). The electrostatic behavior is
captured in Fig.\ \ref{fig1}(c), where the device is represented by
a capacitance model with tunnel coupling between neighboring dots.
Negligible direct capacitive coupling between QD1 and QD3 is
expected for the serial geometry. Furthermore, cross capacitances
are inserted as deduced from measurements below and capacitive
coupling to the backgate is included.

In the sequential tunneling regime with very high barriers,
transport through the triple quantum dot is only allowed at certain
quadruple points in the three dimensional gate space spanned by
$V_{Tg1}$,$V_{Tg2}$ and $V_{Tg3}$, where four charge states are
degenerate\cite{SchroerPRB}. However, we observe large cotunneling
current in a wide region of gate space, which can be interpreted as
transport through molecular states of the triple quantum dot. The
interdot tunnel couplings ($t_{12}$ and $t_{23}$) formed under the
top-gates are strong enough to form a coherent superposition
(molecular state) over the triple dot, which is weakly coupled to
the Ti source/drain electrodes.

\begin{figure}[t!]
\begin{center}
\includegraphics[width=0.48\textwidth]{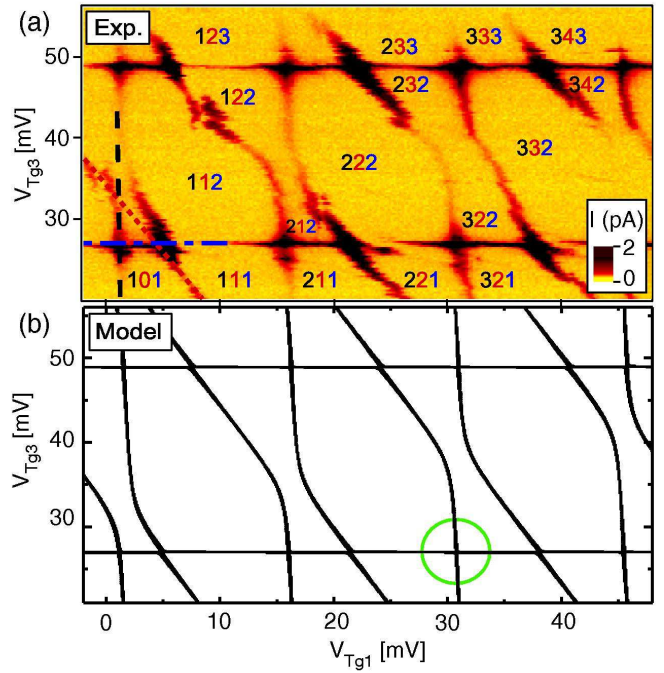}
\end{center}
\caption{(a) Experiment: Current versus $V_{Tg1}$ and $V_{Tg3}$
probing the three quantum dots in the triple quantum dot
($T=50$\,mK, $V_{sd}=50$\,$\mu$V, $V_{bg}=4$\,V, $V_{Tg2}=0$\,V).
Overall vertical, sloping and horizontal lines are observed
representing charge degeneracies between two charge states related
to addition of an electron in QD1, QD2 and QD3, respectively. Strong
and weak anti-crossings are seen between QD1-2 and QD2-3,
respectively. Charge states are indicated by additional electron
number in the quantum dots. (b) Model: Stability diagram based on
the capacitor model in Fig.\ \ref{fig1}(c) including tunnel coupling
between neighboring dots capturing the main features of the
measurement in (a). See also Supporting Information.
\label{figure2} }
\end{figure}
Figure \ref{figure2}(a) shows the current through the triple quantum
dot versus Tg1 and Tg3 with constant voltage on Tg2 at finite bias
voltage having (overall) vertical, sloping and horizontal lines.
Examples are shown by black (dashed), red (dotted) and blue
(dash-dotted) lines in the lower left of the figure\footnote{Several
gate switches were observed during measurements despite keeping the
gate ranges very small. Figure \ref{figure2} has therefore been
translated along the gate axes $\Delta V_{Tg1}= $-6.5\,mV and
$\Delta V_{Tg2}=-1.5$\,V to correct such switches relative to Fig.\
3.}. The vertical, sloping and horizontal lines correspond to adding
an electron to QD1, QD2 and QD3, respectively. It is seen that
sweeping only Tg1 adds electrons to QD1 (crossing vertical lines)
and QD2 (crossing sloping lines), while sweeping only Tg3 adds
electron to QD2 (crossing sloping lines) and QD3 (crossing
horizontal lines). Tg1/3 therefore couples capacitively to both
QD1/3 and QD2 as schematically drawn in Fig.\ \ref{fig1}(c). The
charge states $N_1N_2N_3$ are shown in the plot, where $N_i$ is the
additional electron number in QD $i$, $i=1,2,3$. The observed
behavior can, however, not solely be explained by electrostatics
(without coupling). A relative large tunnel and capacitive coupling
is seen at (anti)crossings between sloping and vertical lines in
contrast to the crossings between sloping and horizontal lines. Thus
the capacitive ($C_{m12}$) and tunnel coupling ($t_{12}$) between
QD1 and QD2 are relatively large compared to the capacitive
$C_{m23}$ and tunnel coupling $t_{23}$ between QD2 and QD3 (DQD23).
Furthermore, the direct capacitive coupling between QD1 and QD3 is
as expected small because a crossing behavior is observed between
horizontal and vertical lines. Since an electron can not tunnel from
QD1 to QD3, no direct tunnel coupling between QD1 and QD3 exists in
contrast to the triangular triple quantum dot geometry. This
crossing will be subject for further discussion below.

\begin{figure}[t!]
\begin{center}
\includegraphics[width=0.48\textwidth]{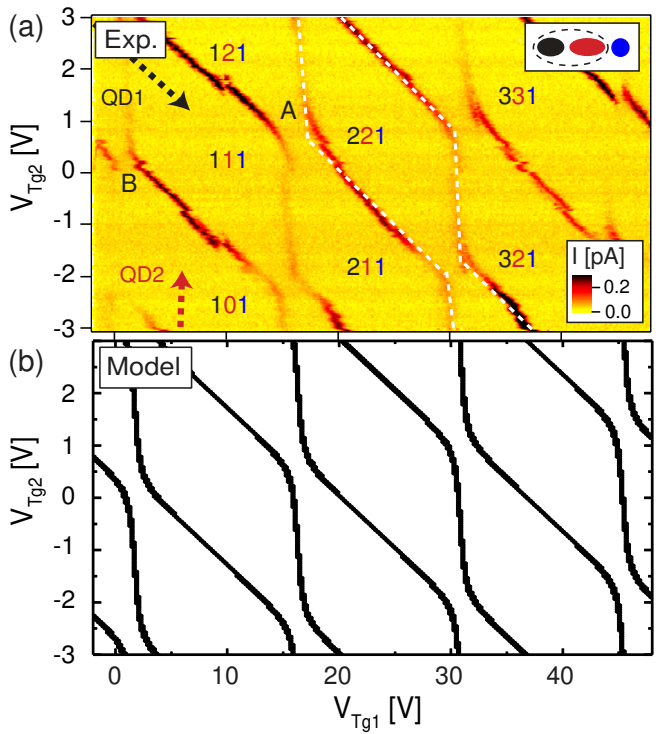}
\end{center}
\caption{Experiment: Current versus $V_{Tg1}$ and $V_{Tg2}$ at
$V_{bg}=4$\,V, $V_{sd}= 50$\,$\mu$V and T=50\,mK of the double
quantum dot formed by QD1 and QD2 as shown in the inset. QD3 is held
in Coulomb blockade ($V_{Tg3}=16$\,mV). A clear (tilted) honeycomb
pattern indicated by white dashed lines is seen showing two strongly
coupled quantum dots and revealing a significant (negligible) cross
capacitance from Tg1 to QD2 (Tg2 to QD1). The three colored numbers
indicate the charge state of additional electron numbers in each QD.
The red and black dotted arrows show the gate voltages for the
single quantum dot measurements in Fig.\ \ref{fig4}(a-b). (b) Model:
Calculation of the stability diagram of the double quantum dot
within the triple quantum dot showing good agreement with the
experiment ($V^{model}_{Tg3}=16$\,mV, {\em i.e.}, Coulomb blockade).
See Supporting Information for more details. \label{fig3}}
\end{figure}
We can also probe a more familiar charging effect on a double
quantum dot formed by QD1 and QD2 (DQD12).
Figure \ref{fig3}(a) shows the current through the triple quantum
dot versus Tg1 and Tg2 at finite bias voltage. The voltage on Tg3 is
fixed leaving QD3 in Coulomb blockade with a constant number of
electrons. A (tilted) honeycomb pattern is revealed as expected for
a double quantum dot with finite current primarily along sloping and
vertical lines due to electrostatics. Crossing the vertical lines
correspond to adding an electron to QD1, while crossing the sloping
lines similarly correspond to adding an electron to QD2. When
increasing Tg1 for constant Tg2 both vertical and sloping lines are
crossed indicating a capacitive coupling from Tg1 to both QD1 and
QD2, respectively, as deduced above as well (Fig.\ \ref{figure2}).
In contrast only sloping lines are crossed when increasing Tg2 for
constant Tg1 illustrating that there is no cross capacitance from
Tg2 to QD1 (thus part of the deduced schematics in Fig.\
\ref{fig1}(c)). Consistent with Fig.\ \ref{figure2}(a), the strong
anti-crossings observed between sloping and horizontal lines reveals
the presence of a significant electrostatic ($C_{m12}$) and tunnel
coupling ($t_{12}$) between QD1 and QD2. This indicates that
transport involves molecular ground states, {\em e.g.}, the
transport around A (lower wing) is due to a degeneracy between the
111 charge state and the ground state of the quantum superposition
involving the 121 and 211 charge states. Some variation in the
tunnel and capacitive coupling is observed, {\em e.g.}, stronger
coupling between the 211 and the 121 states (anti-crossing at A)
than between the 101 and 011 states (anti-crossing at B).

\begin{figure}[t!]
\begin{center}
\includegraphics[width=0.44\textwidth]{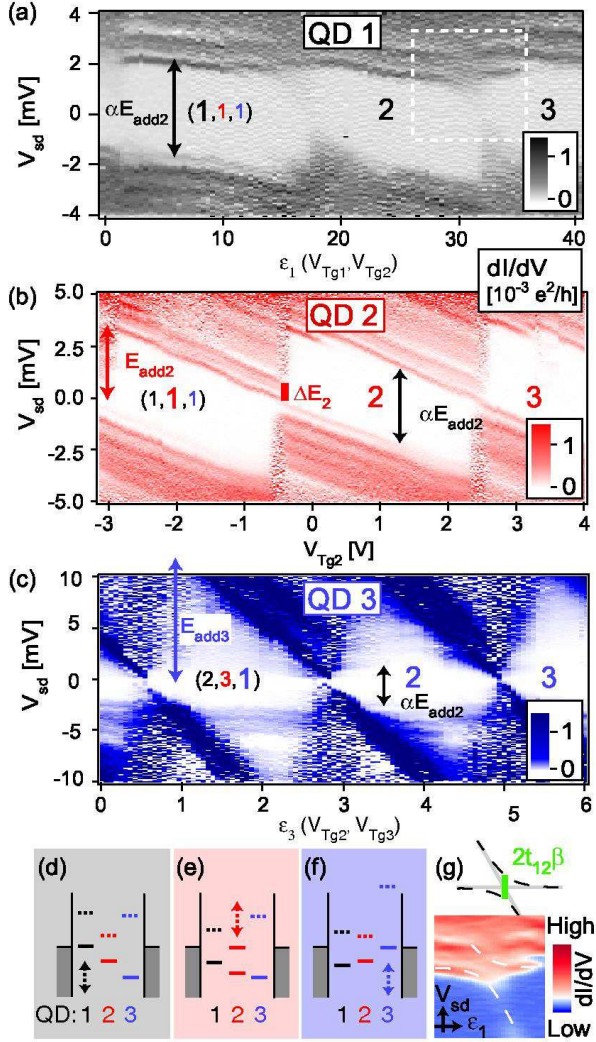}
\end{center}
\caption{(a-c) Stability diagrams of QD 1-3 (blue, red and black dot
in Fig.\ \ref{fig1}(b,c)) showing that three single quantum dots are
formed at $V_{bg}=4$\,V ($T=50$\,mK). The additional electron number
in each dot is indicated in the center of the Coulomb blockade
diamonds. Red and blue arrows indicate the addition energy of QD2
and QD3, while black arrows show the condition for finite bias
transport through molecular states mostly belonging to QD2 in
Coulomb blockade for QD1 and QD3 ($\alpha$ is a capacitance
dependent factor). (d-f) Energy diagrams of the triple quantum dot
showing the electrochemical potentials for adding electrons to the three
dots. When measuring the stability diagram of {\em e.g.}\ QD1 (d),
the electrochemical potentials in QD2 and QD3 are held constant, while
tuning the electrochemical potentials in QD1 up or down (indicated by
dotted arrows). Similarly, QD2 and QD3 ((e) and (f)) can be probed.
(g) Data from dashed square in QD1 (a) showing level crossing
between levels in QD1 and QD2, i.e., the Coulomb blockade diamond
edge (mainly level in QD1) anti-crosses the horizontal lines (mainly
level in QD2) (see schematics). Dashed and gray lines illustrate the
behavior with and without coupling, respectively.
 \label{fig4}}
\end{figure}
We can also investigate each single QD by sweeping the top-gate
voltages, for example along the black or red dashed arrows in Fig.\
\ref{fig3}(a) for the measurement of QD1 and QD2, respectively.
Similarly,  QD3 can be probed as well (not shown)\footnote{Condition
for top-gate voltages Tg1-3, QD1: $V_{Tg1}=1$\,mV
$\times\epsilon_1-2$\,mV, $V_{Tg2}=-0.15$\,V $\times\epsilon_1+3$\,V
($N_2=1$), $V_{Tg3}=16$ mV ($N_3=1$); QD2: $V_{Tg1}=7$\,mV
($N_1=1$), $V_{Tg3}=16$\,mV ($N_3=1$); QD3: $V_{Tg1}=29$\,mV
($N_1=2$), $V_{Tg2}=1$\,V $\times\epsilon_3-2$\,V ($N_2=3$),
$V_{Tg3}=10.5$\,mV $\times \epsilon_3-3$\,mV.}. Figure \ref{fig4}(b)
shows the stability diagram of QD2 by adjusting the electrochemical
potentials of QD2 only (see Fig.\ \ref{fig4}(e)) via Tg2. It reveals
clear Coulomb blockade diamonds due to addition of electrons in QD2
with an addition energy of $E_{add2} \sim 3.5$\,meV and excited
states with a level spacing of $\Delta E_2 \sim 0.7$ meV (see arrow
and vertical bar, respectively). Similarly, the stability diagram of
QD3 in Fig.\ \ref{fig4}(c) shows well defined Coulomb blockade
diamonds with an estimated addition energy of $E_{add3}\sim 12$\,meV
(blue arrow) and a less clear level spacing around $\Delta
E_3\sim1-2$\,meV. However, additional horizontal lines are visible
at low finite bias attributed to transport through molecular states
primarily related to the electron being localized in QD2 ({\em i.e.}
source or drain chemical potentials being aligned with (red)
electrochemical potentials of QD2 in Fig.\ \ref{fig4}(f)). The bias gap
between the highest negative and lowest positive lines therefore
corresponds to the addition energy of QD2 scaled by a capacitance
dependent factor $\alpha$ indicated by the black arrow in Fig.\
\ref{fig4}(a-c). Finally, the stability diagram of QD1 is shown in
Fig.\ \ref{fig4}(a) exhibiting a less regular behavior. The Coulomb
diamonds are cut off at finite bias due to tunneling through
molecular states as the case of QD3, which makes it more difficult
to estimate the addition energy. An estimate of diamond (2,1,1)
yields $E_{add1}\sim 6$\,meV. Also in this case the scaled addition
energy of QD2 matches the bias gap (black arrow) between lowest
finite bias lines, and the fine structure at higher bias is in
agreement with the level spacing of QD2. The bias voltage position
of the finite bias lines in QD1 (Fig.\ \ref{fig4}(a)) are shifted by
changing voltage on Tg2, i.e., tuning the electrochemical potential
(molecular states) of QD2 consistent with the above interpretation
(not shown).

It should be noted that the transport characteristics is different
from weakly coupled triple quantum dots in the sequential tunneling
regime. Transport is allowed even when only one of the QD is located
in the transport window while others are in the Coulomb blockade
regime. We observe larger current when two QDs are in the transport
window as seen at the crossing points of conductance lines in Fig.\
\ref{figure2}, and even larger current at the quadruple point (not
shown). A more thorough discussion of the current in terms of
cotunneling follows below. We also note that the observed
characteristic of relatively strong interdot tunnel coupling is
incompatible with transport through three parallel CNTs, since they
only would be capacitively coupled. The molecular state
interpretation of the triple dot measurements is further supported
by the following calculation based on the capacitance model shown in
Fig.\ \ref{fig1}(c). We try to estimate a reasonable set of
parameters (12 capacitances and two tunnel couplings) by calculating
stability diagrams of the capacitor circuit shown in Fig.\
\ref{fig1}(c) including single electron tunneling processes within
the triple quantum dot. The general features of the resulting
stability diagram (the vertical, sloping and horizontal lines)
are due to electrostatics, which are primarily determined by the
top-gate capacitances \cite{vanderWiel,SchroerPRB} (see Supporting
Information S1). The deduced charging energies $U_{c1}$,$U_{c2}$ and
$U_{c3}$ from the single quantum dot measurements in Fig.\
\ref{fig4}(a-c) depend on the total capacitance of each dot and are
related to the capacitances of the model by ($C_{m12},C_{m23} \ll
C_{1,2,3}$)\cite{SchroerPRB}
\begin{eqnarray}
 U_{c1} & \simeq & \frac{e^2}{C_1}(1-\frac{C_{m23}^2}{C_2 C_3}) \\
  U_{c2} & \simeq & \frac{e^2}{C_2}\\
 U_{c3} & \simeq & \frac{e^2}{C_3}(1-\frac{C_{m12}^2}{C_1 C_2})
\end{eqnarray}
where $C_i$, $i=1,2,3$ is the sum of the capacitances directly
connected to QD $i$, {\em i.e.}, $C_1=C_s+C_{Tg1}+C_{bg1}+C_{m12}$,
$C_2=C_{m12}+C_{Tg2}+C_{bg2}+C_{m23}+C_{Tg1QD2}+C_{Tg3QD2}$ and
$C_3=C_d+C_{Tg3}+C_{bg3}+C_{m23}$ (see Supporting Information S1).
The charging energies obtained and used in the model are $U_{c1}
\simeq 6.0$\,meV,$U_{c2} \simeq 3.5$\,meV and $U_{c3} \simeq
12.0$\,meV. As pointed out in the analysis of the measurements, the
coupling between QD1 and QD2 is stronger than the coupling between
QD2 and QD3 yielding coupling energies $U_{cm12}\simeq
0.26$\,meV$~>U_{cm23} \simeq 0.05$\,meV (see Supporting
Information). Geometrical considerations on the device structure are
also used to estimate the backgate capacitances\footnote{The length
between the contacts and the lengths of QD1-3 are $L \sim
1.4$\,$\mu$m and $L_{QDi}$, $i=1,2,3$ with $L_{QD2}\sim 1$\,$\mu$m
fixed (assuming the CNT crosses perpendicular to the contacts). The
top-gates are made in a second electron beam lithography step
aligning to predefined markers and we here assume a 100\,nm shift
(often observed) to the right relative to the contacts making QD3
small consistent with measurements (high charging energy $U_{c3}$).
The estimated lengths of QD1 and QD2 therefore become $L_{QD1}\sim
0.3$\,$\mu$m and $L_{QD3}\sim 0.1$\,$\mu$m yielding backgate
capacitances according to $C_{bg,i}=2\pi\epsilon L_i / ln(4h/d)$,
where $\epsilon=3.9\epsilon_0$ for Si-oxide with height $h=500$\,nm
and $d\sim 1$\,nm is the diameter of the CNT\cite{McEuen}.} and the
12 capacitances are listed in Table \ref{tabcap}.

Given all the capacitances in the model, electrostatic stability
diagrams can be calculated\cite{SchroerPRB} by finding the charge
(ground) state $(N_1,N_2,N_3)$ having the lowest electrostatic
energy $U_{N_1N_2N_3}$. When the tunnel couplings are included, the
charge states with the same total number of electrons on the triple
quantum dot couples together forming molecular states ({\em e.g.}\
$N=1$, (1,0,0), (0,1,0) and (0,0,1) couple together). The
eigenenergies for a total electron number $N$ on the triple quantum
dot is found by diagonalising the corresponding matrix with the
diagonal elements given by the electrostatic energies
$U_{N_1N_2N_3}$ and the off-diagonal elements being zero except when
two charge states are connected by one tunneling event, {\em e.g.},
(1,0,0) is connected with (0,1,0) by $t_{12}$ and (0,1,0) is
connected with (0,0,1) by $t_{23}$. When the molecular ground state
energy with a total number of $N$ electrons on the triple quantum
dot is equal to the ground state energy with $N+1$ electrons,
transport is allowed. The model therefore describes the measurement
by ground state transport through molecular triple quantum dot
states. More details are given in the Supporting Information S2.
\begin{table}
\begin{center}
\begin{tabular}{c|c||c|c||c|c||c|c}
  $i$    & $C_i$ [aF]  & $i$  &  $C_{i}$ [aF] & $i$    & $C_i$ [aF]  & $i$  &  $C_{i}$ [aF] \\
  \hline
  Tg1    & 11         &   bg1    &  8.6  &  Tg1QD2   &    9.5         &  cm12     &    2    \\
  Tg2    & 0.05       &  bg2    & 28.5  &  Tg3QD2    &    6           &  cm23        &    0.2   \\
  Tg3    &  7.3       &   bg3    &  2.9  &  s        &    2.9           &  d        &     5  \\
 \end{tabular}
\end{center}
\caption{Table showing the capacitances used in the model to
calculate the stability diagrams in Fig.\ \ref{figure2}(b) and Fig.\
\ref{fig3}(b) from which the charging/coupling energies are
calculated. Note, the low capacitance value of Tg2, an effect also
observed in other devices with this geometry, probably due to some
damage during the processing.}\label{tabcap}
\end{table}

Figure \ref{figure2}(b) shows the resulting plot, which resembles
the main features of the measurement in Fig.\ \ref{figure2}(a) with
good agreement\footnote{The origin (0,0,0) of
$(V_{Tg1},V_{Tg2},V_{Tg3})$ in the model is translated to
(-5.5\,mV,0\,V,16\,mV) to match the experimental voltages exactly in
Fig.\ \ref{figure2} and \ref{fig3}. Furthermore, the model
calculation in Fig.\ \ref{fig3} has been shifted by the addition of
one electron in QD2, i.e., $-\Delta V_{Tg2}=-3.019$\,V, since the
model triple quantum dot is empty for negative voltages on
$V_{Tg2}$.}. The electrostatic part of the model captures the
charging effects, while the tunnel part (and capacitive interdot
coupling) makes strong/weak anti-crossing behavior for DQD12/DQD23.
We obtain tunnel couplings $t_{12} \sim 0.3$\,meV and $t_{23}
\lesssim 0.03$\,meV from the analysis, where the tunnel coupling
$t_{23}$ is an upper bound. An estimate of the anti-crossing
 between levels in QD1 and QD2 ($t_{12}$) can also be found from the
bias stability diagram of QD1 shown in Fig.\ \ref{fig4}(a), e.g., in
the dashed box shown in another color scheme in Fig.\ \ref{fig4}(g).
The lower white dashed line in Fig.\ \ref{fig4}(g) follows the edge
of the Coulomb diamond (mainly a level in QD1), which makes an
anti-crossing with a horizontal line (mainly a level in QD2) as the
bias is increased. See also Fig.\ \ref{fig4}(a), where no dashed
guide lines are drawn. The schematic of this behavior is shown in
the top panel of Fig.\ \ref{fig4}(g) with the strength of the
anti-crossing being pure tunnel-like ({\em i.e.} no interdot
capacitance). An estimate along the bias direction (green bar)
yields $t_{12}\sim 0.35$\,meV in agreement with the average value
obtained from the model. The capacitance dependent factor $\beta$ is
in this case close to 1, since the electrochemical potentials in QD1 are
only little affected by changing the voltage on the source. A more
complicated anti-crossing behavior involving excited states is also
observed at higher bias voltage. Finally, the model also describes
the measurements when sweeping another set of top-gates as {\em
e.g.} shown in Fig.\ \ref{fig3}(a-b) for DQD12. Since Tg1 and Tg2 do
not couple to QD3, a well coupled double quantum dot honeycomb
structure is observed consistent with the triple quantum dot
stability diagram. A double quantum dot DQD23 is also formed between
QD2 and QD3, which shows very weak anti-crossings as expected (not
shown) from the above measurement.

The model presented here is the simplest approach to a coupled
triple quantum dot ignoring shell structure, spin and the effect of
hybridization to the leads. The shell structure is to some extent
visible in the measurement, since the addition energies are not
constant throughout the gate sweeps shown. This is supported by the
single QD stability diagrams, where excited states are observed
(Fig.\ \ref{fig4}). However, the shell structure is unfortunately
not clear enough to assign even/odd or four-fold electron occupation
for the filling. These effects are therefore not included in the
model in the simplest approach. In case of clear shell structure, an
interesting topic would be to study exchange interaction in the
triple quantum dot system. The model assumes constant (average)
tunnel and capacitive couplings even though the coupling between the
different charge states (orbitals) vary in the measurements as
mentioned.

\begin{figure}[t!]
\begin{center}
\includegraphics[width=0.48\textwidth]{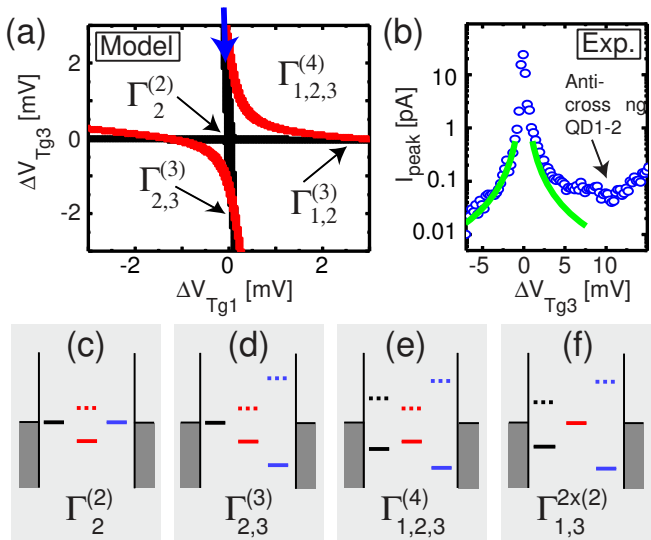}
\end{center}
\caption{(a) Model: The crossing (black lines) and 2nd order
anti-crossing (red lines) correspond to having the electrochemical
potential aligned in QD1 and QD3. Tunnel couplings
$t_{12}=t_{23}=0.5$\,meV are chosen large to increase clarity.
Examples of gate-regions with 2nd (c), 3rd (d) and 4th (e) order
cotunneling rates $\Gamma^{(i)}_j$ with $(i)$ being the order and
$j$ referring to the QD(s) in Coulomb blockade. (b) Measured peak
current along the vertical (and for large positive $\Delta V_{Tg3}$
sloping) resonance line extracted from Fig.\ \ref{figure2}(a) with
the peak position marked by the green circle in Fig.\
\ref{figure2}(b). The green lines are best fit based on 3rd order
cotunneling to the tails of the peak. (c-e) Energy diagrams showing
the above mentioned cotunneling configurations. (f) Another possible
cotunneling process involving two consecutive 2nd order processes.
 \label{fig5}}
\end{figure}

We will end by briefly discussing two phenomena in triple quantum
dots based on the above model and experiment. We focus on a gate
voltage-region, where a vertical and a horizontal line cross as {\em
e.g.} marked by the green circle in the model calculation of Fig.\
\ref{figure2}(b) and magnified in Fig.\ \ref{fig5}(a) (black lines).
The electrochemical potential of QD1 and QD3 at the crossing are aligned
with the chemical potential of the leads, while QD2 is in Coulomb
blockade as shown in Fig.\ \ref{fig5}(c). The current profile along
the vertical {\em resonance} line indicated by the blue arrow in
Fig.\ \ref{fig5}(a) is therefore expected to be a single peak. This
is confirmed in Fig.\ \ref{fig5}(b) showing the measured peak
current profile (blue circles) versus voltage on Tg3 extracted from
Fig.\ \ref{figure2}(a), where $\Delta V_{Tg3}=0$ is at the crossing.
However, increasing the tunnel coupling (mainly $t_{23}$) in the
model leads to a clear anti-crossing (red lines in Fig.\
\ref{fig5}(a)) even though no direct tunnel coupling between QD1 and
QD3 exists. Such an anti-crossing is therefore of second order, a
phenomenon not possible to study in double quantum dots. The system
has close analogy to a three-level system in the
$\Lambda$-configuration, where intriguing experiments are expected
\cite{BrandesReview}.
Although the anti-crossing is not resolved in our device, second or
higher order cotunneling can be discussed in terms of elastic
cotunneling rates $\Gamma^{(i)}_j$, where $(i)$ is the order and $j$
is the QD(s) in Coulomb blockade. In the close vicinity of the above
discussed crossing, only QD2 is in Coulomb blockade (see Fig.\
\ref{fig5}(c)), and the major contribution to the current therefore
stems from second order cotunneling with rate $\Gamma^{(2)}_2$,
yielding a relatively high current. Detuning the electrochemical potential
of QD1 (QD3) along the horizontal (vertical) resonance line changes
the order of the dominant cotunneling to third order $\Gamma^{(3)}$
(black arrows in Fig.\ \ref{fig5}(a)). The electrons now have to
cotunnel through two neighboring QDs, {\em i.e.}, QD2 and QD1 (QD3)
as seen in Fig.\ \ref{fig5}(d) for $\Gamma^{(3)}_{2,3}$. The current
therefore decreases as illustrated by the peak in Fig.\
\ref{fig5}(b). The green lines show best fit (using the data for
negative $\Delta V_{Tg3}$) to the measurement
$I\sim\Gamma^{(3)}_{2,3} \sim 1/\Delta V_{Tg3}^2$, which is the
expected gate-dependence sufficiently far from resonance
\cite{Flensberg}. Here the third order process is dominant and we
assume negligible contribution due to the simultaneous change of the
electrochemical potential of QD2, when changing $V_{Tg3}$. A less good
correspondence between the fit and the measurement is observed on
the right side of the peak, which might be related to the strong
anti-crossing between QD1 and QD2 for positive $\Delta V_{Tg3}$
modifying the gate-dependence. For gate voltages outside the
 lines (white areas in Fig.\ \ref{fig5}(a)), all QDs are in Coulomb
blockade (Fig.\ \ref{fig5}(e)). Current is then due to fourth order
cotunneling $\Gamma^{(4)}_{1,2,3}$ and therefore highly suppressed
giving the stable charge configurations in the stability diagrams of
Fig.\ \ref{figure2}(a) and Fig.\ \ref{fig3}(a). For completeness,
one additional type of process exists involving two second order
cotunneling $\Gamma^{2x(2)}_{1,3}$ corresponding to having only the
electrochemical potential of QD2 at resonance (see Fig.\ \ref{fig5}(f)).
This process is more likely than third order cotunneling
\cite{SchroerPRB} consistent with the higher current along sloping
lines ($\Gamma^{2x(2)}_{1,3}$) than vertical lines
($\Gamma^{(3)}_{2,3}$) in the double quantum dot (DQD12) stability
diagram shown in Fig.\ \ref{fig3}(a).

In conclusion we have shown measurement on a top-gated single wall
carbon nanotube interpreted as a serially coupled triple quantum dot
formed between the source and drain electrodes. The stability
diagram for single, double and triple quantum dot(s) are observed by
individual control of the electron number on each quantum dot via
top-gate voltages. An electrostatic model describing ground state
transport involving triple quantum dot molecular states captures the
main features of the double and triple quantum dot stability
diagrams with good agreement. Finally, second order anti-crossings
and cotunneling was discussed.

{\bf Acknowledgement:} We would like to thank Y. Hirayama for
support of this joint collaboration and A. Fujiwara and Y. Ono for
use of their equipment. Furthermore, we like to acknowledge the
support of the EU-STREP Ultra-1D and CARDEQ programs. This work was
also supported by the SCOPE from the Ministry of Internal Affairs
and Communications of Japan, and by a Grant-in-Aid for Scientific
Research from the JSPS.

{\bf Supporting Information available:} The derivation of the
electrostatic energy used in the model is explained in details
together with the procedure to obtain stability diagrams without
tunnel coupling (S1). Furthermore, the method to include tunnel
coupling between neighboring quantum dots yielding the stability
diagrams of the triple quantum dot shown in this Letter is addressed
(S2).



\end{document}